# Gridless Quadrature Compressive Sampling with Interpolated Array Technique

Feng Xi [1], Shengyao Chen [1], Yimin D. Zhang [2], Zhong Liu [1]

[1] Department of Electronic Engineering
Nanjing University of Science and Technology
Nanjing, Jiangsu 210094, People's Republic of China
Email: eezliu@mail.njust.edu.cn

[2] Center for Advanced Communications
Villanova University, Villanova, PA 19085, USA

*Abstract*—Quadrature compressive sampling (QuadCS) is a sub-Nyquist sampling scheme for acquiring in-phase and quadrature (I/Q) components in radar. In this scheme, the received intermediate frequency (IF) signals are expressed as a linear combination of time-delayed and scaled replicas of the transmitted waveforms. For sparse IF signals on discrete grids of time-delay space, the QuadCS can efficiently reconstruct the I/Q components from sub-Nyquist samples. In practice, the signals are characterized by a set of unknown time-delay parameters in a continuous space. Then conventional sparse signal reconstruction will deteriorate the QuadCS reconstruction performance. This paper focuses on the reconstruction of the I/Q components with continuous delay parameters. A parametric spectrum-matched dictionary is defined, which sparsely describes the IF signals in the frequency domain by delay parameters and gain coefficients, and the QuadCS system is reexamined under the new dictionary. With the inherent structure of the QuadCS system, it is found that the estimation of delay parameters can be decoupled from that of sparse gain coefficients, yielding a beamspace direction-of-arrival (DOA) estimation formulation with a time-varying beamforming matrix. Then an interpolated beamspace DOA method is developed to perform the DOA estimation. An optimal interpolated array is established and sufficient conditions to guarantee the successful estimation of the delay parameters are derived. With the estimated delays, the gain coefficients can be conveniently determined by solving a linear least-squares problem. Extensive simulations demonstrate the superior performance of the proposed algorithm in reconstructing the sparse signals with continuous delay parameters.

## I. INTRODUCTION

Quadrature sampling [1,2] is a commonly used processing scheme in radar systems to obtain the baseband *in-phase* and *quadrature* (respectively denoted as *I* and *Q*) components. For the intermediate frequency (IF) signals with center frequency $f_c$ and bandwidth $B$, the quadrature sampling theorem [1] states that digital *I* and *Q* components can be acquired with a sampling frequency

$$f_{IF} = \frac{4f_L + 2B}{4d+1}, \qquad (1)$$

where $d$ is a positive integer satisfying $d \leq \lfloor f_L/(2B) \rfloor$ and $f_L = f_c - B/2$, where $\lfloor \cdot \rfloor$ denotes a floor function resulting in the largest integer not exceeding the argument. With appropriate setting of $f_c$, the minimum sampling rate $2B$ can be allocated. The requirement (1) has become a serious bottleneck in the development of wideband/ultrawideband systems. Inspired by the compressive sampling (CS) [3-5] and analog-to-information (A2I) conversion [6-11], we recently developed a quadrature compressive sampling (QuadCS) system [12,13] for the radar echo signals consisting of a linear combination of the time-delayed and scaled replicas of the transmitted waveforms. The QuadCS scheme collects compressive *I* and *Q* components at a sub-Nyquist rate and permits perfect reconstruction of these components at the Nyquist rate. Its advantages have been shown in performance analyses [14] and application in pulse-Doppler processing [15]. The QuadCS system is also applicable to channel estimation in communication [16] and navigation [17] systems in which the received signal is characterized by a multipath environment. In this paper, we focus our description on radar applications.

The accurate reconstruction capability of the QuadCS system relies on the knowledge of the sparsifying dictionary. For applications to radar [13-15], the waveform-matched dictionary [18] is often assumed, *i.e.*, the dictionary consists of a finite set of time-delayed versions of the known transmitting waveform at the Nyquist-sampling grids along the time delay axis. When the delays of the targets are exactly on the discrete grids, the dictionary well represents the radar echo signal. However, in many practical scenarios, the delays of the targets, or other parameters that characterize the sparse signals, are continuous and, in general, do not lie exactly on the pre-defined discrete grids. In this case, the reconstruction performance degrades significantly because of the mismatch between the assumed dictionary and the practical signals [13]. This problem is referred to as the off-grid problem in CS theory [19].

The off-grid problem has attracted significant attention over the past few years and some solutions have been reported, e.g., [20-30]. These techniques can be classified into four categories. The first one [20, 21] is to discretize the delay axis or parameter space with finer grids when applying CS techniques. The finer grids will result in highly coherent dictionaries. The second one [22, 23] is to model the perturbation caused by the grid mismatch as Gaussian or uniformly distributed noise and the sparse Bayesian interference [31,32] is then used to solve the off-grid problem. Instead of discretizing the parameter space, methods in the third category [24-27] jointly optimize the recovered signals and the dictionary to alleviate the effect of the basis mismatch. Although these methods may reduce the reconstruction error to some extent, they require a high computation complexity and often lead to numerical instability. The last category [28-30] is to use the atom-norm proposed in [33] to handle the infinite dictionary defined on the continuous parameter. Although the atom-norm based technique is promising, a big challenge is to effectively solve the atom-norm minimization problem. Currently, its application is limited to the line spectral estimation in Fourier basis [28-30], where the minimization is cast into a convex semidefinite programing problem. It remains unclear how to generalize it to other scenarios.

In this paper, we focus on the signal reconstruction for the QuadCS system with sparse parameters defined in a continuous parameter space. We first define a parametric spectrum-matched dictionary, which well describes the received radar signals in frequency domain by delay parameters and gain coefficients. Then, by exploiting the inherent structure of the QuadCS system and realigning the QuadCS compressive measurements, we find that the estimation of the delay parameters can be equivalently formulated as the direction-of-arrival (DOA) estimation problem in array processing [34,35]. The DOA estimation techniques can then be utilized to estimate the delay parameters. With the estimated delay parameters, the estimation of the gain coefficients is performed by conventional least-squares technique. Under the proposed reconstruction scheme, discretization of the continuous parameter space is no longer necessary. We decompose the joint estimation of the delay parameters and the gain coefficients into two separate estimation problems. As such, both computational complexity and numerical stability are greatly improved.

Different from conventional DOA estimation formulations, the proposed approach estimates DOAs (which represents delay parameters) in the beamspace with a time-varying beamforming matrix, *i.e.*, the equivalent beamspace array data are generated by a set of time-varying beamformers operating on the array outputs. Therefore, while a number of beamspace DOA estimation methods [36-38] exist, the time-varying beamforming matrix prevents their direct application to the underlying problem. To solve this problem, we develop an interpolated beamspace DOA method, which stems from the interpolated array techniques [39-41], for the estimation of delay parameters. Optimal array interpolations are established and the sufficient conditions are derived to guarantee the successful estimation of the delay parameters. Extensive simulations demonstrate that the proposed algorithm can achieve super-resolution time-delay estimation and high-accuracy sparse signal reconstruction.

Note that the idea of using DOA estimation for sparse time-delay estimation was used in [42], where a sub-Nyquist sampling system was proposed under the framework of the union of subspace (UoS) [43]. Through multi-channel sampling, the time-delay estimation is directly transformed into DOA estimation with a uniform linear array. Our works differ from [42] in the research scope and the developed DOA estimation model. As to be discussed in Section IV, the DOA estimation model is defined by realigning single-channel samples and is established in beamspace with time-varying beamformers. The lack of existing beamspace DOA estimation techniques for such situations motivated us to develop an interpolated array based approach to successfully solve the underlying problem.

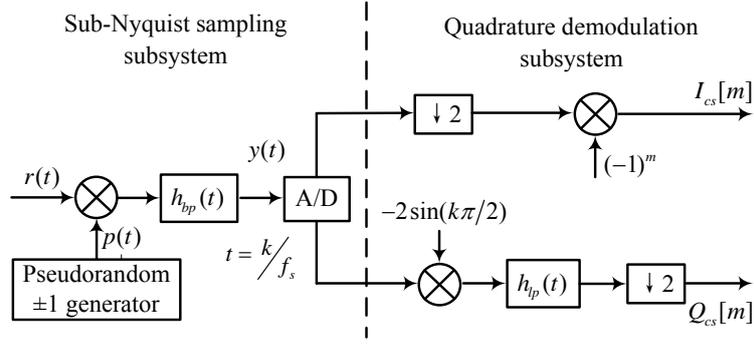

Fig.1 The structure of the QuadCS system.

The rest of the paper is organized as follows. The signal model and problem formulation are given in Section II. The QuadCS system is reexamined from the frequency-domain point of view in Section III. The DOA-based time-delay estimation is developed in Section IV. The scheme of the gridless signal reconstruction is given in Section V. In section VI, optimal design on interpolated array and beamforming matrix is established. Section VII shows simulation results and Section VII concludes the paper.

*Notations*: Bold-face letters are reserved for vectors and matrices. $(\cdot)^*$, $(\cdot)^T$ and $(\cdot)^H$ denote complex conjugation, transposition, and conjugate transposition, respectively. $\|\cdot\|_1$ and $\|\cdot\|_2$ denote the $l_1$-norm and $l_2$-norm, respectively. $\text{Re}\{\cdot\}$ and $\text{Im}\{\cdot\}$ respectively represent the real part and imaginary part of a complex argument. To avoid confusion, a real-valued signal, a complex signal, and the Fourier transform of a signal are represented as $s(t)$, $\tilde{s}(t)$ and $\hat{S}(f)$, respectively. $\lfloor \cdot \rfloor$ and $\lceil \cdot \rceil$ respectively denotes the floor and the ceiling functions, and $\text{diag}(\mathbf{x})$ represents a diagonal matrix with the elements of vector $\mathbf{x}$ as its diagonal elements. In addition, $\text{tr}(\cdot)$ denotes a matrix trace.

## II. SIGNAL MODEL AND PROBLEM FORMULATION

We consider an IF system in which the received signal $x(t)$ is expressed as

$$x(t) = \sum_{k=1}^{K} v_k a(t - \tau_k) \cos\left[2\pi f_0 (t - \tau_k) + \phi(t - \tau_k) + \varphi_k\right], \quad (2)$$

where $a(t)$ and $\phi(t)$ are the envelope and phase of *a priori* known waveform, respectively, $f_0$ is the IF frequency, and $\tau_k$, $v_k$ and $\varphi_k$ are the unknown parameters of the time delay, gain coefficient and phase offset of the $k$-th target, respectively, for $k = 1, \cdots, K$. The signal model (2) is generic and is widely used in radar systems.

Denote $\tilde{s}_0(t) = a(t) e^{j\phi(t)}$ as the complex baseband signal and $\tilde{v}_k = v_k e^{j\varphi'_k}$ as the complex gain coefficient with $\varphi'_k = \varphi_k - 2\pi f_0 \tau_k$. The complex envelope $\tilde{s}(t)$ of the IF signal $x(t)$ is given by

$$\tilde{s}(t) = \sum_{k=1}^{K} \tilde{v}_k \tilde{s}_0(t - \tau_k). \quad (3)$$

The *I* and *Q* components of the IF signal $x(t)$ are respectively the real and imaginary parts of the complex envelope $\tilde{s}(t)$, expressed as

$$I(t) = \operatorname{Re}\{\tilde{s}(t)\} = \sum_{k=1}^{K} v_k a(t - \tau_k) \cos\left[\phi(t - \tau_k) + \varphi'_k\right], \quad (4)$$

$$Q(t) = \operatorname{Im}\{\tilde{s}(t)\} = \sum_{k=1}^{K} v_k a(t - \tau_k) \sin\left[\phi(t - \tau_k) + \varphi'_k\right]. \quad (5)$$

We assumes that the IF signal $x(t)$ has a bandpass spectrum with bandwidth $B$ centered at frequency $f_0$ ($f_0 \gg B/2$). Let $\hat{S}_0(f)$ be the Fourier transform of $\tilde{s}_0(t)$. Then, the Fourier transform $\hat{S}(f)$ of $\tilde{s}(t)$, expressed below, is bandlimited to $[-B/2, B/2)$,

$$\hat{S}(f) = \int_{-\infty}^{\infty} \tilde{s}(t) e^{-j2\pi f t} dt = \sum_{k=1}^{K} \tilde{v}_k \hat{S}_0(f) e^{-j2\pi f \tau_k}. \quad (6)$$

With the known IF frequency $f_0$, the useful information of the IF signal $x(t)$ is completely characterized by the complex envelope $\tilde{s}(t)$ or, equivalently, the *I* and *Q* components $I(t)$ and $Q(t)$. The aim of the digital quadrature demodulation is to acquire the digital *I* and *Q* components from the IF signal $x(t)$. Note that both $\tilde{s}(t)$ as well as or $I(t)$ and $Q(t)$ are uniquely defined by the delays $\tau_k$ and the complex gains $\tilde{v}_k$, $k=1,\cdots,K$. As such, the problem is equivalent to determining these parameters from the samples of $x(t)$.

In [13], we developed a QuadCS system to acquire the complex envelope $\tilde{s}(t)$, in which the delays $\tau_k$ are assumed to lie in the discrete grids of the delay space $(0,\tau_{max}]$ and the $\tilde{s}(t)$ is sparsely represented by the waveform-matched dictionary. In this paper, we assume that the delays $\tau_k$ are continuous in the delay space $(0,\tau_{max}]$. To efficiently represent the signal $\tilde{s}(t)$, we define a parametric dictionary in the frequency domain, $\{\hat{\psi}(f,\tau)|\hat{\psi}(f,\tau)=\hat{S}_0(f)e^{-j2\pi f\tau}, 0<\tau\leq\tau_{max}\}$. It is seen that, for a continuous parameter $\tau$, the dictionary consists of infinite number of atoms $\hat{\psi}(f,\tau)$, which are the phase-modulated replicas of the spectrum $\hat{S}_0(f)$. For convenience, we refer to the dictionary as parametric *spectrum-matched* dictionary. With this dictionary, the spectrum of $\tilde{s}(t)$ in (6) can be represented as

$$\hat{S}(f)=\sum_{k=1}^{K}\tilde{v}_k\hat{\psi}(f,\tau_k)=\hat{\boldsymbol{\psi}}(f,\boldsymbol{\tau})\tilde{\mathbf{v}}, \qquad(7)$$

where $\boldsymbol{\tau}=\{\tau_1,\tau_2,\cdots,\tau_K\}$, $\tilde{\mathbf{v}}=[\tilde{v}_1,\tilde{v}_2,\cdots,\tilde{v}_K]^T$, and

$$\hat{\boldsymbol{\psi}}(f,\boldsymbol{\tau})=[\hat{\psi}(f,\tau_1),\hat{\psi}(f,\tau_2),\cdots,\hat{\psi}(f,\tau_K)]. \qquad(8)$$

For any given parameter set $\boldsymbol{\tau}$, $\hat{\boldsymbol{\psi}}(f,\boldsymbol{\tau})$ determines a $K$-dimensional subspace. Then the acquisition of $\hat{S}(f)$, or equivalently $\tilde{s}(t)$, is to optimize the delay parameter set $\boldsymbol{\tau}$ along with the complex gain vector $\tilde{\mathbf{v}}$ such that the parametric dictionary approaches the true subspace in which $\hat{S}(f)$ lies. In CS theory, the problem is called as joint dictionary learning and signal recovery.

III. QUADRATURE COMPRESSIVE SAMPLING IN FREQUENCY DOMAIN

Now we introduce the QuadCS system that performs sub-Nyquist sampling of the received IF signals as expressed in (2). Specifically, we examine the QuadCS system shown in Fig. 1 from the frequency-domain point of view and formulate the signal reconstruction problem under the framework of parametric spectrum-matched dictionary.

*A. Sampling Scheme*

The QuadCS system in Fig. 1 consists of two parts: the low-rate sampling subsystem and the quadrature demodulation subsystem. The former performs the mixing, bandpass filtering, and the bandpass sampling to yield low-rate output sampling sequence, whereas the latter implements the digital quadrature demodulation [1] to generate the compressive I and Q components.

More specifically, in the low-rate sampling subsystem, the IF signal $x(t)$ is first modulated by a random spectrum-spreading signal $p(t)$, which is $T_p$-periodic, i.e.,

$$p(t) = \sum_{l=-L_p}^{L_p} \tilde{\rho}_l e^{j2\pi f_p l t}, \qquad (9)$$

where $f_p = 1/T_p$, $\tilde{\rho}_l$ is the Fourier series of $p(t)$, and $L_p \geq BT_p$ is a positive integer. It is implicitly implied that the highest frequency of $p(t)$ is no less than $B$. The mixing operation spreads the frequency spectrum of the resulting baseband signal to span the entire full spectrum of $p(t)$. The spectrum of the mixed output is given as

$$\hat{X}_p(f) = \sum_{l=-L_p}^{L_p} \tilde{\rho}_l \hat{X}(f - f_p l). \qquad (10)$$

After mixing, the signal spectrum $\hat{X}_p(f)$ is truncated by a bandpass filter $h_{bp}(t)$ with the bandwidth $B_{cs} \ll B$ centered at $f_0$. Assume that the bandpass filter is ideal and has the frequency response as

$$\hat{H}_{bp}(f) = \begin{cases} B/B_{cs}, & f \in \mathcal{B}_{cs}, \\ 0, & \text{elsewhere}, \end{cases} \qquad (11)$$

where $\mathcal{B}_{cs} = (-f_0 - B_{cs}/2, -f_0 + B_{cs}/2] \cup [f_0 - B_{cs}/2, f_0 + B_{cs}/2)$. Then, the spectrum of the bandpass filter output $y(t)$ is

$$\hat{Y}(f) = B/B_{cs} \sum_{l=-L_0}^{L_0} \tilde{\rho}_l \hat{X}(f - f_p l), \qquad f \in \mathcal{B}_{cs}, \tag{12}$$

where $L_0$ is determined by

$$L_0 = \left\lceil \frac{B + B_{cs}}{2 f_p} \right\rceil - 1. \tag{13}$$

Let the output $y(t)$ be sampled according to the bandpass sampling theorem with the sampling frequency $f_s = (4 f_{cs}^L + 2 B_{cs})/(4d+1)$ in which $f_{cs}^L = f_0 - B_{cs}/2$ and $d$ is a positive integer satisfying $d \leq \lfloor f_{cs}^L / 2 B_{cs} \rfloor$. To simplify the analysis, we assume that the minimum bandpass sampling rate $f_s = 2 B_{cs}$ is achieved. Then, the spectrum of discretely sampled $y(t)$ can be expressed as

$$\begin{aligned} \hat{Y}(e^{j 2\pi f/f_s}) &= 2 B_{cs} \sum_{k=-\infty}^{+\infty} \hat{Y}(f - f_s k) \\ &= 2B \sum_{k=-\infty}^{+\infty} \sum_{l=-L_0}^{L_0} \tilde{\rho}_l \hat{X}(f - f_s k - f_p l) \\ &= B \sum_{l=-L_0}^{L_0} \tilde{\rho}_l \sum_{k=-\infty}^{+\infty} \left( \hat{S}(f - B_{cs}/2 - f_s k - f_p l) + \hat{S}^*(f + B_{cs}/2 - f_s k - f_p l) \right). \end{aligned} \tag{14}$$

The quadrature demodulation subsystem performs similarly as in the classic digital quadrature demodulation [1] to extract compressive I and Q components ($I_{cs}[m]$ and $Q_{cs}[m]$) from the output of the bandpass sampling. With the operations of the quadrature demodulation, the spectra of $I_{cs}[m]$ and $Q_{cs}[m]$ can be derived as

$$\hat{I}_{cs}(e^{j 2\pi f/B_{cs}}) = \frac{B}{2} \sum_{l=-L_0}^{L_0} \tilde{\rho}_l \sum_{k=-\infty}^{+\infty} \left( \hat{S}(f - B_{cs} k - f_p l) + \hat{S}^*(f - B_{cs} k - f_p l) \right), \tag{15}$$

$$\hat{Q}_{cs}(e^{j 2\pi f/B_{cs}}) = \frac{B}{2} j \sum_{l=-L_0}^{L_0} \tilde{\rho}_l \sum_{k=-\infty}^{+\infty} \left( \hat{S}^*(f - B_{cs} k - f_p l) - \hat{S}(f - B_{cs} k - f_p l) \right). \tag{16}$$

Let $S_{cs}[m] = I_{cs}[m] + j Q_{cs}[m]$. Then the spectrum of $S_{cs}[m]$ is given by

$$\hat{S}_{cs}(e^{j 2\pi f/B_{cs}}) = B \sum_{l=-L_0}^{L_0} \tilde{\rho}_l \sum_{k=-\infty}^{+\infty} \hat{S}(f - B_{cs} k - f_p l). \tag{17}$$

Since $\hat{S}_{cs}(e^{j 2\pi f/B_{cs}})$ is $B_{cs}$-periodic, we can confine our analysis in one period $[-B_{cs}/2, B_{cs}/2)$, i.e.,

$$\hat{S}_{cs}\left(e^{j2\pi f/B_{cs}}\right) = B\sum_{l=-L_0}^{L_0} \tilde{\rho}_l \hat{S}(f - f_p l), \qquad f \in [-B_{cs}/2, B_{cs}/2]. \tag{18}$$

It is shown that the spectrum $\hat{S}_{cs}\left(e^{j2\pi f/B_{cs}}\right)$ is the lowpass-filtered output of a linear combination of $2L_0+1$ frequency-shifted versions of $\hat{S}(f)$.

*B. Signal Reconstruction under Parametric Spectrum-Matched Dictionary*

The signal reconstruction is equivalent to reconstructing the spectrum $\hat{S}(f)$ of the complex envelope $\tilde{s}(t)$ from the output CS spectrum $\hat{S}_{cs}\left(e^{j2\pi f/B_{cs}}\right)$. Referring to (7), we can re-express $\hat{S}_{cs}\left(e^{j2\pi f/B_{cs}}\right)$ in (18) as

$$\hat{S}_{cs}\left(e^{j2\pi f/B_{cs}}\right) = B\sum_{l=-L_0}^{L_0} \tilde{\rho}_l \sum_{k=1}^{K} \tilde{v}_k \hat{\psi}(f - f_p l, \tau_k) = \sum_{k=1}^{K} \tilde{v}_k \hat{\phi}(f, \tau_k), \tag{19}$$

where

$$\hat{\phi}(f, \tau) = B\sum_{l=-L_0}^{L_0} \tilde{\rho}_l \hat{\psi}(f - f_p l, \tau). \tag{20}$$

We show in (19) that the output spectrum $\hat{S}_{cs}\left(e^{j2\pi f/B_{cs}}\right)$ can also be represented by a set of $K$ atoms $\hat{\phi}(f,\tau)$ defined by the parameter set $\{\tau_1, \tau_2, \cdots, \tau_K\}$.

In practice, we can only obtain a finite-length discrete spectrum of $\hat{S}_{cs}\left(e^{j2\pi f/B_{cs}}\right)$. Let $\Delta f = B_{cs}/L$ be the Nyquist sampling interval in frequency-domain, where $L > 0$ is the length of the sampling sequence[1]. The discrete sampling $\hat{S}_{cs}[l]$ of the $\hat{S}_{cs}\left(e^{j2\pi f/B_{cs}}\right)$ is given as

$$\hat{S}_{cs}[l] = \hat{S}_{cs}\left(e^{j2\pi f_{(l)}/B_{cs}}\right) \tag{21}$$

with $f_{(l)} = -B_{cs}/2 + (l-1)\Delta f$ ($l = 1, 2, \cdots, L$). Then we can rewrite (19) in the following matrix form,

$$\hat{\mathbf{s}}_{cs} = \hat{\mathbf{\Phi}}(\boldsymbol{\tau})\tilde{\mathbf{v}}, \tag{22}$$

where

$$\hat{\mathbf{s}}_{cs} = \left[\hat{S}_{cs}[1], \hat{S}_{cs}[2], \cdots, \hat{S}_{cs}[L]\right]^T \in \mathbb{C}^L, \tag{23}$$

---

[1] With this setting, we implicitly assume that the signal $x(t)$ has finite support with length $L/B_{cs}$.

$$\hat{\boldsymbol{\Phi}}(\boldsymbol{\tau}) = \begin{bmatrix} \hat{\phi}(f_{(1)}, \tau_1) & \hat{\phi}(f_{(1)}, \tau_2) & \cdots & \hat{\phi}(f_{(1)}, \tau_K) \\ \hat{\phi}(f_{(2)}, \tau_1) & \hat{\phi}(f_{(2)}, \tau_2) & \cdots & \hat{\phi}(f_{(2)}, \tau_K) \\ \vdots & \vdots & \ddots & \vdots \\ \hat{\phi}(f_{(L)}, \tau_1) & \hat{\phi}(f_{(L)}, \tau_2) & \cdots & \hat{\phi}(f_{(L)}, \tau_K) \end{bmatrix} \in \mathbb{C}^{L \times K}. \tag{24}$$

In this sense, the signal reconstruction problem is a joint estimation of the delay parameter set $\boldsymbol{\tau}$ and the gain parameter vector $\tilde{\mathbf{v}}$ from the frequency-domain sampling vector $\hat{\mathbf{s}}_{cs}$ such that the vector $\hat{\mathbf{s}}_{cs}$ can be represented by atoms as few as possible. In the LASSO formulation [44], it is equivalent to solve

$$\min_{\boldsymbol{\tau}, \tilde{\mathbf{v}}} \frac{1}{2} \left\| \hat{\mathbf{s}}_{cs} - \hat{\boldsymbol{\Phi}}(\boldsymbol{\tau}) \tilde{\mathbf{v}} \right\|_2^2 + \gamma \left\| \tilde{\mathbf{v}} \right\|_1, \tag{25}$$

where $\gamma > 0$ is the regularization parameter balancing the least-square constraint and the sparsity constraint. Different from traditional sparse reconstruction problems, the formulation in (25) is a joint optimization of the dictionary and the sparse signal. Several methods [24-27] have been developed to solve such a joint optimization problem. However, they are computationally expensive and are very likely to be trapped in the undesirable local minima.

## IV. THE GRIDLESS TIME-DELAY PARAMETERS ESTIMATION

In this section, we show that the joint estimation of the delay parameter set $\boldsymbol{\tau}$ and the gain parameter vector $\tilde{\mathbf{v}}$ can be decomposed and the delay estimation can be performed by DOA estimation based techniques. Sufficient conditions to guarantee the successful estimation of the unknown delays are derived.

### A. *Delay Parameter Estimation vs. Direction of Arrival Estimation*

To show the relation between delay estimation and DOA estimation, we first reformulate the frequency-domain sampling vector of the QuadCS system. Let $M = B_{cs}/f_p$ and $N = L/M$ be some positive integer. We extract $N$ down-sampling sequences $\hat{\mathbf{s}}_{cs}^{[n]}$ ($n = 1, 2, \cdots, N$) with down-sampling rate $N$ from the sequence $\hat{S}_{cs}[l]$, i.e.,

$$\hat{\mathbf{s}}_{cs}^{[n]} = \left[ \hat{S}_{cs}[n], \hat{S}_{cs}[N+n], \hat{S}_{cs}[(M-1)N+n] \right]^T \in \mathbb{C}^M. \tag{26}$$

Then we can decompose (22) as $N$ measurement equations in the frequency domain,

$$\hat{\mathbf{s}}_{cs}^{[n]} = \hat{\mathbf{\Phi}}^{[n]}(\boldsymbol{\tau})\tilde{\mathbf{v}}, \quad n = 1, 2, \cdots, N \tag{27}$$

where

$$\hat{\mathbf{\Phi}}^{[n]}(\boldsymbol{\tau}) = \begin{bmatrix} \hat{\phi}(f_{(n)}, \tau_1) & \cdots & \hat{\phi}(f_{(n)}, \tau_K) \\ \hat{\phi}(f_{(n+N)}, \tau_1) & \cdots & \hat{\phi}(f_{(n+N)}, \tau_K) \\ \vdots & \ddots & \vdots \\ \hat{\phi}(f_{(n+(M-1)N)}, \tau_1) & \cdots & \hat{\phi}(f_{(n+(M-1)N)}, \tau_K) \end{bmatrix} \in \mathbb{C}^{M \times K}. \tag{28}$$

By (20), the $(k,m)$-th element of the matrix $\hat{\mathbf{\Phi}}^{[n]}(\boldsymbol{\tau})$ is given by

$$\begin{aligned}
\hat{\phi}(f_{(n+(m-1)N)}, \tau_k) &= B \sum_{l=-L_0}^{L_0} \tilde{\rho}_l \hat{\psi}(f_{(n+(m-1)N)} - lf_P, \tau_k) \\
&= B \sum_{l=-L_0}^{L_0} \tilde{\rho}_l \hat{\psi}(f_{(n)} + (m-1)f_p - lf_P, \tau_k) \\
&= B \sum_{l'=-L_0-m+1}^{L_0-m+1} \tilde{\rho}_{l'+m-1} \hat{\psi}(f_{(n)} - l'f_p, \tau_k) \\
&= B \sum_{l'=-L_0-m+1}^{L_0-m+1} \tilde{\rho}_{l'+m-1} \hat{S}_0(f_{(n)} - l'f_p) e^{j2\pi(l'f_p - f_{(n)})\tau_k},
\end{aligned} \tag{29}$$

where the last equality follows the definition of the parametric spectrum-matched dictionary in Section II.

With (29), we can factorize the matrix $\hat{\mathbf{\Phi}}^{[n]}(\boldsymbol{\tau})$. To simplify the derivation, assume $M = 2M_0$ to be even and define $J \triangleq 2L_0 + 1 - M = 2(L_0 - M_0) + 1$. Inserting (29) into (28), we have the following factorization,

$$\hat{\mathbf{\Phi}}^{[n]}(\boldsymbol{\tau}) = \mathbf{P}\hat{\mathbf{S}}^{[n]}\mathbf{W}(\boldsymbol{\tau})\mathbf{D}^{[n]}(\boldsymbol{\tau}), \tag{30}$$

where the matrices $\mathbf{P} \in \mathbb{C}^{M \times J}$ and $\mathbf{W}(\boldsymbol{\tau}) \in \mathbb{C}^{J \times K}$ are given as

$$\mathbf{P} = B \times \begin{bmatrix} \tilde{\rho}_{-L_0} & \tilde{\rho}_{-L_0+1} & \cdots & \tilde{\rho}_{L_0-M} \\ \tilde{\rho}_{-L_0+1} & \tilde{\rho}_{-L_0+2} & \cdots & \tilde{\rho}_{L_0-M+1} \\ \vdots & \vdots & \ddots & \vdots \\ \tilde{\rho}_{-L_0+M-1} & \tilde{\rho}_{-L_0+M} & \cdots & \tilde{\rho}_{L_0-1} \end{bmatrix}, \tag{31}$$

$$\mathbf{W}(\boldsymbol{\tau}) = \begin{bmatrix} 1 & 1 & \cdots & 1 \\ e^{j2\pi f_p \tau_1} & e^{j2\pi f_p \tau_2} & \cdots & e^{j2\pi f_p \tau_K} \\ \vdots & \vdots & \ddots & \vdots \\ e^{j2\pi(J-1)f_p \tau_1} & e^{j2\pi(J-1)f_p \tau_2} & \cdots & e^{j2\pi(J-1)f_p \tau_K} \end{bmatrix}, \tag{32}$$

and $\hat{\mathbf{S}}^{[n]} \in \mathbb{C}^{J \times J}$ and $\mathbf{D}^{[n]}(\boldsymbol{\tau}) \in \mathbb{C}^{K \times K}$ are diagonal matrices defined as

$$\hat{\mathbf{S}}^{[n]} = \mathrm{diag}\left(\left[\hat{S}_0\left(f_{(n)} + L_0 f_p\right), \hat{S}_0\left(f_{(n)} + (L_0-1)f_p\right), \cdots, \hat{S}_0\left(f_{(n)} - (L_0-M)f_p\right)\right]\right), \tag{33}$$

$$\mathbf{D}^{[n]}(\boldsymbol{\tau}) = \mathrm{diag}\left(\left[e^{j2\pi f_p(-(L_0-M_0)-(n-1)/N)\tau_1}, e^{j2\pi f_p(-(L_0-M_0)-(n-1)/N)\tau_2}, \cdots, e^{j2\pi f_p(-(L_0-M_0)-(n-1)/N)\tau_K}\right]\right). \tag{34}$$

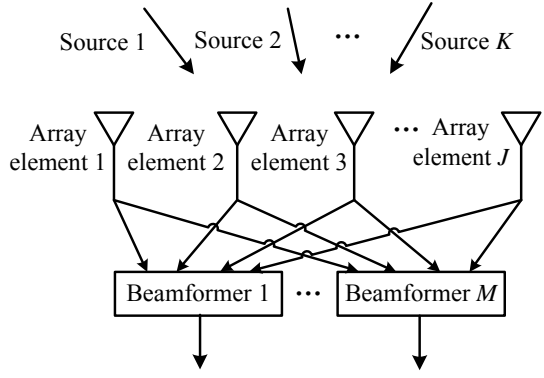

Fig.2 The array structure in beamspace DOA estimation.

Note that the matrix $\mathbf{P} \in \mathbb{C}^{M \times J}$ is a partial Toeplitz matrix [45] and $\mathbf{W}(\boldsymbol{\tau})$ is a Vandermonde matrix.

Define $\mathbf{B}^{[n]} = \mathbf{P}\hat{\mathbf{S}}^{[n]}$, $\tilde{\mathbf{v}}^{[n]} = \mathbf{D}^{[n]}(\boldsymbol{\tau})\tilde{\mathbf{v}}$, and $\boldsymbol{\theta} = \{\theta_1, \theta_2, \cdots, \theta_K\}$, where $\theta_k = 2\pi f_p \tau_k$ ($1 \leq k \leq K$). Then, (27) can be rewritten as

$$\hat{\mathbf{s}}_{cs}^{[n]} = \mathbf{B}^{[n]} \mathbf{W}(\boldsymbol{\theta}) \tilde{\mathbf{v}}^{[n]}, \quad n = 1, 2, \cdots, N, \tag{35}$$

where

$$\mathbf{W}(\boldsymbol{\theta}) = \begin{bmatrix} 1 & 1 & \cdots & 1 \\ e^{j\theta_1} & e^{j\theta_2} & \cdots & e^{j\theta_K} \\ \vdots & \vdots & \ddots & \vdots \\ e^{j(J-1)\theta_1} & e^{j(J-1)\theta_2} & \cdots & e^{j(J-1)\theta_K} \end{bmatrix}. \tag{36}$$

The formulation (35) resembles exactly that of DOA estimation in beamspace [36-38]. Then the delay parameters and gain coefficients can be estimated separately. In (35), the array is uniformly linear with $J$ array elements; $\mathbf{W}(\boldsymbol{\theta})$ is the array manifold whose columns are the steering vectors towards the directions $\theta_1, \theta_2, \cdots, \theta_K$, respectively; $\tilde{\mathbf{v}}^{[n]}$ is the source signal vector with $k$-th signal from the direction $\theta_k$; $\mathbf{W}(\boldsymbol{\theta})\tilde{\mathbf{v}}^{[n]}$ is the array data vector at the $n$-th snapshot; and $\mathbf{B}^{[n]}$ is a beamforming matrix which formulates $M$ beamformers. Under the framework, we obtain $N$ beamspace data vectors $\hat{\mathbf{s}}_{cs}^{[n]}$ ($n = 1, 2, \cdots, N$) which are used to extract the DOA information. The array structure of the formulated DOA problem is shown in Fig. 2.

*B. Delay Parameters Estimation*

The beamspace DOA estimation has received considerable attention in array signal processing and a variety of beamspace DOA estimation methods [36-38] have been developed. However, different from the conventional formulations in [36-38], the beamforming matrix in (35) is time-varying. Thus, conventional beamspace DOA estimation methods cannot be directly applied to solve this problem.

In this section, we resort to the interpolated array techniques to develop a beamspace DOA estimation with time-varying beamforming matrix. The basic idea is to design a set of interpolated matrices for each snapshot and transform the time-varying beamforming matrix into a time-invariant one so that conventional beamspace DOA estimation methods can be applied.

Denote $\mathbf{a}(\theta) \in \mathbb{C}^M$ as the targeted steering vector towards the direction $\theta$ which takes account of the effects of array structure and beamformers. The interpolation is to construct $N$ interpolated matrices $\mathbf{T}^{[n]} \in \mathbb{C}^{M \times M}, n = 1, 2, \cdots, N$, such that

$$\mathbf{T}^{[n]}\mathbf{B}^{[n]}\mathbf{w}(\theta) = \mathbf{a}(\theta) \tag{37}$$

for $\theta \in \Theta$, where $\Theta \subset (0, 2\pi]$ is a sector containing the set of unknown DOAs. In the design of interpolated array, we require that the array manifold after interpolation has the structure similar to that before interpolation to minimize the interpolation error [46, 47]. Toward this end, one of the possible choices of the targeted steering vector is

$$\mathbf{a}(\theta) = \mathbf{B}_0 \mathbf{w}(\theta), \tag{38}$$

where $\mathbf{B}_0 \in \mathbb{C}^{M \times J}$ is a time-invariant beamforming matrix. The design on the beamforming matrix is critical for high-performance DOA estimation and will be discussed in details in Section VI.

With the $N$ interpolated matrices, we obtain output vectors $\bar{\mathbf{s}}_{cs}^{[n]}$ of the interpolated array

$$\bar{\mathbf{s}}_{cs}^{[n]} = \mathbf{T}^{[n]} \hat{\mathbf{s}}_{cs}^{[n]} = \mathbf{A}(\boldsymbol{\theta}) \tilde{\mathbf{v}}^{[n]}, \quad n = 1, 2, \cdots, N, \tag{39}$$

where

$$\mathbf{A}(\boldsymbol{\theta}) = \left[ \mathbf{a}(\theta_1), \mathbf{a}(\theta_2), \cdots, \mathbf{a}(\theta_K) \right] \in \mathbb{C}^{M \times K}. \tag{40}$$

The correlation matrix of the interpolated array $\bar{\mathbf{R}}_{cs}$ can be formed as

$$\bar{\mathbf{R}}_{cs} = \frac{1}{N} \sum_{n=1}^{N} \bar{\mathbf{s}}_{cs}^{[n]} \left( \bar{\mathbf{s}}_{cs}^{[n]} \right)^H = \mathbf{A}(\boldsymbol{\theta}) \tilde{\mathbf{R}}_v \mathbf{A}^H(\boldsymbol{\theta}), \tag{41}$$

where

$$\tilde{\mathbf{R}}_v = \frac{1}{N} \sum_{n=1}^{N} \tilde{\mathbf{v}}^{[n]} \left( \tilde{\mathbf{v}}^{[n]} \right)^H. \tag{42}$$

Then eigen decomposition techniques can be used to determine the unknown directions $\boldsymbol{\theta}$ of the source signals. Let us perform an eigen-decomposition of the matrix $\bar{\mathbf{R}}_{cs}$,

$$\bar{\mathbf{R}}_{cs} = [\mathbf{E}, \mathbf{G}] \begin{bmatrix} \boldsymbol{\Lambda}_s & 0 \\ 0 & 0 \end{bmatrix} \begin{bmatrix} \mathbf{E}^H \\ \mathbf{G}^H \end{bmatrix}, \tag{43}$$

where $\boldsymbol{\Lambda}_s = \text{diag}(\sigma_1^2, \sigma_2^2, \cdots, \sigma_K^2)$ with the $K$ non-zero eigenvalues of $\bar{\mathbf{R}}_{cs}$, $\sigma_1^2, \sigma_2^2, \cdots, \sigma_K^2$, in non-increasing order; $\mathbf{E} \in \mathbb{C}^{M \times K}$ consists of the eigenvectors corresponding to the $K$ non-zero eigenvalues of $\bar{\mathbf{R}}_{cs}$ and $\mathbf{G} \in \mathbb{C}^{M \times (M-K)}$ consists of the eigenvectors corresponding to the remaining $M-K$ eigenvalues of $\bar{\mathbf{R}}_{cs}$. The $K$ source signals belong to the signal subspace $\text{span}(\mathbf{E})$, which is orthogonal to the noise subspace $\text{span}(\mathbf{G})$. With the eigen-decomposition, various existing subspace-based algorithms DOA estimation methods can be applied. In this paper, we take the well-known beamspace MUSIC [36] as an example to perform the simulation analyses in Section VII.

The set of DOAs $\boldsymbol{\theta}$ is estimated by searching the $K$ peaks of the MUSIC pseudo-spectrum, expressed as

$$P(\theta) = \frac{1}{\mathbf{a}^H(\theta)\mathbf{G}\mathbf{G}^H\mathbf{a}(\theta)}. \qquad (44)$$

For the estimated set of DOAs, $\bar{\boldsymbol{\theta}} = \{\bar{\theta}_1, \bar{\theta}_2, \cdots, \bar{\theta}_K\}$, the unknown delays can be calculated as

$$\bar{\tau}_k = \bar{\theta}_k / 2\pi f_p \qquad k = 1, 2, \cdots, K \qquad (45)$$

**Remark 1**: In (37), we assume that the sector $\Theta \subset (0, 2\pi]$ containing the set of unknown DOAs is known in advance. In interpolated array processing, the sector is often determined with the prior information of the signal arrivals or preliminary estimates by using other DOA estimation methods [46, 47]. Moreover, while we have considered $\Theta$ as a consecutive sector, it can also be the union of several sub-sectors. As pointed in [13], the CS-based signal reconstruction with discrete time-delays can locate the actual time-delays in an interval with high hit-rate. For this reason, we may use the CS-based signal reconstruction [13] to form the interpolation sector $\Theta$. Denoting the set of preliminary estimates of the DOAs as $\{\theta'_1, \theta'_2, \cdots, \theta'_K\}$, we can form the sector $\Theta$ as $\Theta = \Theta_1 \cup \Theta_2 \cup \cdots \cup \Theta_K$, where each sub-sector $\Theta_k = [\theta'_k - \Delta_k, \theta'_k + \Delta_k]$ with $\Delta_k$ as the predefined interval.

**Remark 2:** Note that the DOA $\theta$ is $2\pi$-periodic. Then by (45), there exist ambiguous estimates of the delay parameters if $\tau > 1/f_p$, i.e., the largest unambiguous delay is $1/f_p$. For the input signals with the given delay space $(0, \tau_{max}]$, we can set the parameter $f_p$ to satisfy $f_p \leq 1/\tau_{max}$ to ensure unambiguous estimation.

*C. Sufficient Condition for Time-Delay Parameters Estimation*

In the above development, we have implicitly assumed that the source correlation matrix $\tilde{\mathbf{R}}_v$ is of full rank, i.e., $\text{rank}(\tilde{\mathbf{R}}_v) = K$, and the interpolated array can uniquely localize the $K$ sources. Here we give sufficient conditions for the unique estimation.

***Proposition 1:*** For arbitrary $K$ distinct delays $\tau_k \in (0, \tau_{\max}]$, $k = 1, 2, \cdots, K$, if $f_p \leq N/\tau_{\max}$ and $N \geq K$, the source correlation matrix $\tilde{\mathbf{R}}_v$ is positive definite, *i.e.*, $\tilde{\mathbf{R}}_v \succ 0$.

***Proof:*** With (42), the source correlation matrix $\tilde{\mathbf{R}}_v$ can be rewritten as

$$\tilde{\mathbf{R}}_v = \frac{1}{N} \mathbf{V}\mathbf{V}^H, \tag{46}$$

where $\mathbf{V} = \left[\tilde{\mathbf{v}}^{(1)}, \tilde{\mathbf{v}}^{(2)}, \cdots, \tilde{\mathbf{v}}^{(N)}\right] \in \mathbb{C}^{K \times N}$. Note that $\text{rank}(\tilde{\mathbf{R}}_v) = \text{rank}(\mathbf{V})$. Then, if $\text{rank}(\mathbf{V}) = K$, $\text{rank}(\mathbf{V}) = K$ follows.

Following $\tilde{\mathbf{v}}^{[n]} = \mathbf{D}^{[n]}(\boldsymbol{\tau})\tilde{\mathbf{v}}$, we have

$$\mathbf{V} = \text{diag}\left([\tilde{v}'_1, \tilde{v}'_2, \cdots, \tilde{v}'_K]\right) \begin{bmatrix} 1 & e^{-j2\pi f_p \tau_1/N} & \cdots & e^{-j2\pi(N-1)f_p \tau_1/N} \\ 1 & e^{-j2\pi f_p \tau_2/N} & \cdots & e^{-j2\pi(N-1)f_p \tau_2/N} \\ \vdots & \vdots & \ddots & \vdots \\ 1 & e^{-j2\pi f_p \tau_K/N} & \cdots & e^{-j2\pi(N-1)f_p \tau_K/N} \end{bmatrix}, \tag{47}$$

where $\tilde{v}'_k = \tilde{v}_k e^{-j2\pi f_p (L_0 - M_0)\tau_k}$, $k = 1, 2, \cdots, K$. Note that the diagonal matrix $\text{diag}([\tilde{v}'_1, \tilde{v}'_2, \cdots, \tilde{v}'_K])$ in the above expression is of rank $K$. In addition, the $K \times N$-dimensional steering matrix at the right-hand side of (47) is the transpose of a Vandermonde matrix, whose rank is $K$ provided that $f_p \tau_{\max}/N \leq 1$ and $N \geq K$. This proves the result. ∎

***Proposition 2:*** For the steering vector given in (38), if $\text{rank}(\mathbf{B}_0) = M$, any set of the steering vectors $\mathbf{a}(\theta_i)$ associated with $M$ distinct DOAs $\theta_i \in (0, 2\pi]$ ($1 \leq i \leq M$) is linearly independent.

***Proof:*** Assume $M$ distinct DOAs $\theta_i \in (0, 2\pi]$ ($1 \leq i \leq M$). The $M$ steering vectors $\mathbf{a}(\theta_i)$ ($1 \leq i \leq M$) form the following matrix:

$$\left[\mathbf{a}(\theta_1), \mathbf{a}(\theta_2), \cdots, \mathbf{a}(\theta_M)\right] = \mathbf{B}_0 \begin{bmatrix} 1 & 1 & \cdots & 1 \\ e^{j\theta_1} & e^{j\theta_2} & \cdots & e^{j\theta_M} \\ \vdots & \vdots & \ddots & \vdots \\ e^{j(J-1)\theta_1} & e^{j(J-1)\theta_2} & \cdots & e^{j(J-1)\theta_M} \end{bmatrix}. \tag{48}$$

Note that the $J \times M$-dimension matrix at the right-hand side of (48) is a Vandermonde matrix. By applying (13) and $M = B_{cs}/f_p$, we can derive that

$$J = 2\left(\left\lceil \frac{B+B_{cs}}{2f_p}\right\rceil - 1\right) + 1 - M$$

$$> 2\left(\left\lceil \frac{B_{cs}}{f_p}\right\rceil - 1\right) + 1 - M \qquad (49)$$

$$= M - 1,$$

which implies $J \geq M$ for an integer $J$. Therefore, the rank of the $J \times M$-dimension Vandermonde matrix at the right-hand side of (48) is equal to $M$. If $\text{rank}(\mathbf{B}_0) = M$, $\text{rank}([\mathbf{a}(\theta_1), \mathbf{a}(\theta_2), \cdots, \mathbf{a}(\theta_M)]) = M$, i.e., the $M$ steering vectors $\mathbf{a}(\theta_i)$, $1 \leq i \leq M$, are linearly independent. ∎

***Proposition 3:*** Under Proposition 1 and Proposition 2, if $M > K$, the DOA estimation problem (39) derives a unique DOA estimation of $K$ distinct DOAs $\bar{\theta}_k \in (0, 2\pi]$, $k = 1, 2, \cdots, K$.

The proof of the proposition directly follows that of Theorem 1 in [48] and is omitted here.

The above three propositions assure that the array formulation in Subsection IV-A can obtain a unique estimation of DOA parameters $\bar{\boldsymbol{\theta}} = \{\bar{\theta}_1, \bar{\theta}_2, \cdots, \bar{\theta}_K\}$. For length-$L$ frequency-domain sampling vector $\hat{\mathbf{s}}_{cs}$, we have $L = MN$, which implies that $L \geq K(K+1)$ by Propositions 1 and 3. In practice, we have $L = TB_{cs} = TB/\Delta$ ($\Delta = B/B_{cs}$) for the signal with the bandwidth $B$ and the finite-length support $T$. We can derive the following facts about the gridless signal reconstruction:

a) The largest number of delay parameters that can be estimated by the proposed method is $K \leq \lceil \sqrt{TB/\Delta} \rceil - 1$;

b) The minimum bandwidth $B_{cs}$ required to recover the sparse signal consisting of $K$ components is $B_{cs} \geq K(K+1)/T$.

## V. THE SCHEME OF GRIDLESS SIGNAL RECONSTRUCTION

We are now in a position to summarize our gridless signal reconstruction (GLSR) algorithm, including the estimation of both the time delays and the complex gains. Algorithm 1 describes the GLSR process in detail.

In the implementation, with the estimated time-delay parameter set $\bar{\tau}$ in the previous section, the estimation of the gain coefficient vector $\tilde{\mathbf{v}}$ from (22) is equivalent to find the least-squares solution of

$$\min_{\tilde{\mathbf{v}}} \frac{1}{2} \left\| \hat{\mathbf{s}}_{cs} - \hat{\mathbf{\Phi}}(\bar{\tau}) \tilde{\mathbf{v}} \right\|_2^2. \tag{50}$$

Different from joint-optimization of delay parameters and gain coefficients in the LASSO formulation (25), the GLSR algorithm separately estimates the delay parameters and the gain coefficients. The gain estimation (50) is a standard least-squares solution and can be efficiently determined. In the time-delay estimation by DOA techniques, we assume that the number $K$ of the signal components is known in advance. In practice, the number of signals in array processing can be estimated by exploiting the Akaike information criterion or the minimum description length criterion [49].

| **Algorithm 1. GLSR Algorithm** |
|---|
| **Input:** $\hat{\mathbf{S}}_{cs}$, $\hat{S}_0(f)$, $p(t)$, $K$, $B$, $B_{cs}$ |
| **Output:** $\hat{S}(f)$ |
| **Steps:** |
| 1) Generate the sub-sequence vectors $\{\hat{\mathbf{S}}_{cs}^{[1]}, \hat{\mathbf{S}}_{cs}^{[2]}, \cdots, \hat{\mathbf{S}}_{cs}^{[N]}\}$ from the frequency-domain sampling vector $\hat{\mathbf{S}}_{cs}$ as (26); |
| 2) Construct the matrices $\mathbf{P}$, $\hat{\mathbf{S}}^{[n]}$ ($n=1,2,\cdots,N$) as (31) and (33) to form the time-varying beamforming matrix $\mathbf{B}^{[n]}$ and formulate the DOA estimation as (35); |
| 3) Design the time-invariant beamforming matrix $\mathbf{B}_0$ and the set of interpolated matrices $\{\mathbf{T}^{[1]}, \mathbf{T}^{[2]}, \cdots, \mathbf{T}^{[N]}\}$ satisfying (37); |
| 4) Obtain the output vector $\bar{\mathbf{s}}_{cs}^{[n]}$ ($n=1,2,\cdots,N$) of the interpolated array as (39) and compute the correlation matrix $\bar{\mathbf{R}}_{cs}$ as (41); |
| 5) Perform the SVD decomposition of $\bar{\mathbf{R}}_{cs}$ as (43) and determine the signal subspace $\mathbf{E}$ and the noise subspace $\mathbf{G}$; |
| 6) Construct the spectrum function $P(\theta)$ as (44), search the $K$ largest peaks of $P(\theta)$ to find the set $\bar{\boldsymbol{\theta}}$ and estimate the delay parameter set $\bar{\tau} = \bar{\boldsymbol{\theta}}/2\pi f_p$; |
| 7) Construct the matrix $\hat{\mathbf{\Phi}}(\bar{\tau})$ as (24) with the estimated $\bar{\tau}$ and recover the gain coefficient vector $\tilde{\mathbf{v}}$ by solving (50); |
| 8) Reconstruct the spectrum $\hat{S}(f)$ according to (7) by using the estimated values of $\tau$ and $\tilde{\mathbf{v}}$. |

## VI. OPTIMAL TIME-INVARIANT BEAMFORMING MATRIX

As discussed in Section IV, the estimation of delay parameters is formulated as a beamspace DOA problem. However, different from the conventional formulations in [36-38], the beamforming matrix in (35) is time-varying. Based on the interpolated array, this section examines the design of time-varying interpolated matrices $\mathbf{T}^{[n]}$ such that the resulting beamformers $\mathbf{T}^{[n]}\mathbf{B}^{[n]}$ are time-invariant, $\mathbf{T}^{[n]}\mathbf{B}^{[n]} = \mathbf{B}_0$, for all $n \in [1, N]$.

Theoretically, we can specify a time-invariant beamforming matrix $\mathbf{B}_0$ for DOA estimation. However, for our problem, we can jointly design $\mathbf{T}^{[n]}$ and $\mathbf{B}_0$, resulting in an optimal time-invariant beamforming matrix $\mathbf{B}_0$. With the requirement (37) and Proposition 2, the joint design is to minimize the interpolated error, *i.e.*,

$$\begin{cases} \min_{\substack{\mathbf{T}^{[1]},\cdots,\mathbf{T}^{[N]} \in \mathbb{C}^{M \times M} \\ \mathbf{B}_0 \in \mathbb{C}^{M \times J}}} \sum_{n=1}^{N} \int_{\theta \in \Theta} \left\| \mathbf{T}^{[n]}\mathbf{B}^{[n]}\mathbf{w}(\theta) - \mathbf{B}_0\mathbf{w}(\theta) \right\|_2^2 d\theta \\ \text{s.t.} \quad \text{rank}(\mathbf{B}_0) = M. \end{cases} \quad (51)$$

The joint optimization is such that the interpolated beamforming matrix $\mathbf{T}^{[n]}\mathbf{B}^{[n]}$ optimally approximates the time-invariant beamforming matrix $\mathbf{B}_0$ in the minimum least-squares sense. Note that $\mathbf{T}^{[n]}$ is a function of $\mathbf{B}_0$. Then we can first find $\mathbf{T}^{[n]}$ for a given $\mathbf{B}_0$ by solving

$$\min_{\mathbf{T}^{[1]},\cdots,\mathbf{T}^{[N]} \in \mathbb{C}^{M \times M}} \sum_{n=1}^{N} \int_{\theta \in \Theta} \left\| \mathbf{T}^{[n]}\mathbf{B}^{[n]}\mathbf{w}(\theta) - \mathbf{B}_0\mathbf{w}(\theta) \right\|_2^2 d\theta. \quad (52)$$

After simple mathematical manipulation, the solution to (52) is obtained as

$$\mathbf{T}^{[n]} = \mathbf{B}_0 \mathbf{C}_{\mathbf{ww}}^{[n]} \left(\mathbf{B}^{[n]}\right)^H \left(\mathbf{B}^{[n]} \mathbf{C}_{\mathbf{ww}}^{[n]} \left(\mathbf{B}^{[n]}\right)^H\right)^{-1}, \quad (53)$$

where

$$\mathbf{C}_{\mathbf{ww}} = \int_{\theta \in \Theta} \mathbf{w}(\theta) \mathbf{w}^H(\theta) d\theta. \quad (54)$$

Inserting (53) into (51), the problem (51) is transformed into

$$\begin{cases} \min_{\mathbf{B}_0 \in \mathbb{C}^{M \times J}} \operatorname{tr}\left(\mathbf{B}_0 \mathbf{C}_N \mathbf{B}_0^H\right) \\ \text{s.t.} \quad \operatorname{rank}(\mathbf{B}_0) = M, \end{cases} \tag{55}$$

where

$$\mathbf{C}_N = \sum_{n=1}^{N} \left( \mathbf{C}_{ww} - \mathbf{C}_{ww} \left(\mathbf{B}^{[n]}\right)^H \left(\mathbf{B}^{[n]} \mathbf{C}_{ww} \left(\mathbf{B}^{[n]}\right)^H\right)^{-1} \mathbf{B}^{[n]} \mathbf{C}_{ww}^H \right). \tag{56}$$

To derive a closed-form solution of (55), we consider the special case of $\mathbf{B}_0 \mathbf{B}_0^H = \mathbf{I}_M$, i.e., orthonormal beamforming matrix. In this case, (55) becomes

$$\begin{cases} \min_{\mathbf{B}_0 \in \mathbb{C}^{M \times J}} \operatorname{tr}\left(\mathbf{B}_0 \mathbf{C}_N \mathbf{B}_0^H\right) \\ \text{s.t.} \quad \mathbf{B}_0 \mathbf{B}_0^H = \mathbf{I}_M. \end{cases} \tag{57}$$

From (57), it is seen that the optimal time-invariant beamforming matrix $\mathbf{B}_0$ is composed of the $M$ eigenvectors of $\mathbf{C}_N$ associated with the $M$ smallest eigenvalues. Let us perform eigen-decomposition of $\mathbf{C}_N$ as,

$$\mathbf{C}_N = [\mathbf{u}_1, \cdots, \mathbf{u}_J] \begin{pmatrix} \lambda_1 & & 0 \\ & \ddots & \\ 0 & & \lambda_J \end{pmatrix} [\mathbf{u}_1, \cdots, \mathbf{u}_J]^H, \tag{58}$$

where $\lambda_1 \geq \lambda_2 \geq \cdots \geq \lambda_J$ are the eigenvalues of $\mathbf{C}_N$ and $\mathbf{u}_i \in \mathbb{C}^J$ is the eigenvector corresponding to $\lambda_i$, $i = 1, 2, \cdots, J$. Then, the optimal time-invariant beamforming matrix $\mathbf{B}_0$ is given as

$$\mathbf{B}_0 = [\mathbf{u}_{J-M+1}, \mathbf{u}_{J-M+2}, \cdots, \mathbf{u}_J]^H. \tag{59}$$

With above development, the minimum interpolation error (51) is the sum of the $M$ smallest eigenvalues of $\mathbf{C}_N$. With the found optimal $\mathbf{B}_0$, the interpolated matrices $\mathbf{T}^{[1]}, \cdots, \mathbf{T}^{[N]}$ can be computed by (53).

## VII. SIMULATION RESULTS

In this section, we examine various aspects of our proposed GLSR algorithm through the extensive simulation experiments. As a comparison, the performance of the OMP-based signal reconstruction [50] is also demonstrated. In the following, "OMP-1" denotes the signal reconstruction based on the discrete grids with the resolution $\Delta \tau = \tau_0$ ($\tau_0 = 1/B$); "OMP-2" denotes the signal reconstruction based on the refined discrete grids with the resolution $\Delta \tau = 0.5\tau_0$; and "GLSR-1" and "GLSR-2" denote the GLSR algorithm with two settings of the sector $\Theta$. In GLSR-1, the sector is $\Theta = \Theta_1 \bigcup \Theta_2 \bigcup \cdots \bigcup \Theta_K$ with the sub-sector $\Theta_k = \left[ 2\pi f_p (\tau_k - \tau_0), 2\pi f_p (\tau_k + \tau_0) \right]$. In GLSR-2, the sub-sector is $\Theta_k = \left[ 2\pi f_p (\bar{\tau}_k^{cs} - 2\tau_0), 2\pi f_p (\bar{\tau}_k^{cs} + 2\tau_0) \right]$, where $\bar{\tau}_k^{cs}$ is the estimate of $\tau_k$ given by the OMP-1 algorithm. Note that the sector $\Theta$ in GLSR-1 is accurate, but GLSR-2 is practical.

In simulation experiments, the IF signal $x(t)$ is a linear combination of $K$ time-delayed versions of the LFM pulsed signal with the bandwidth 50 MHz and the pulse width 10.24 μs. The IF frequency $f_0$ is set between 200 MHz and 300 MHz to achieve the minimal sampling rate. Without special statements, the time-delay $\tau_k$ of the $k$-th components is randomly chosen from the interval $(0, 10.24]$ μs. The real gain coefficient $\nu_k$ and the phase offset $\varphi_k$ are uniformly distributed between $(0,1]$ and $(0, 2\pi]$, respectively. The observation length is twice of the maximum pulse width, *i.e.*, 20.48 μs. Two bandpass filters with bandwidths $B_{cs} = 12.5$ MHz and $B_{cs} = 10$ MHz are selected, which result in one fourth and one fifth of the Nyquist sampling rates, respectively. With the minimum IF sampling rates, two beamspace DOA estimation arrays with $M = 16$ and $M = 12$ are formulated. The corresponding number of resolvable overlapped time-delay components is 15 and 11, respectively.

*A. Performance of the Time-Delay Estimation*

In this experiment, the phase $\varphi_k$ is randomly chosen from a uniform distribution. To isolate the effect of other signal components, the time delay separation between any two components is set to at least $3\tau_0$, and the gain coefficient $\nu_k$ is set to 1 for all components to avoid masking from strong signal components. Two performance indexes, namely, the probability of successful estimation and the relative root-mean-square time-delay estimation error (RRMS-TDE), are used for performance evaluation. A set of estimate $\bar{\boldsymbol{\tau}} = \{\bar{\tau}_1, \bar{\tau}_2, \cdots, \bar{\tau}_K\}$ is declared to be successful if $|\tau_k - \bar{\tau}_k| \leq \tau_0$ holds for all $\bar{\tau}_k$, $k = 1, 2, \cdots, K$. When successful estimation is achieved, we compute the RRMS-TDE, which is defined as

$$\text{RRMS-TDE} = \frac{1}{\tau_0} \sqrt{\frac{1}{K} \sum_{k=1}^{K} (\bar{\tau}_k - \tau_k)^2} . \qquad (60)$$

The probability of successful estimation is shown in Fig. 3 with respect to the number of components $K$. It is clear that GLSR-1 demonstrates the best performance and OMP-2 is superior to OMP-1. GLSR-2 performs like OMP-2 for small $K$. As $K$ increases, the probability of successful estimation of the GLSR algorithms decreases much faster than that of the OMP-based algorithms. The rapid decrease of the probability for large values of $K$ is due to the increase of the interpolated array errors. For the GLSR algorithms, the errors behave like array noise in the equivalent DOA estimation process. Fig. 4 shows the interpolated array errors ($\text{tr}(\mathbf{B}_0 \mathbf{C}_N \mathbf{B}_0^H)$ in (57)) with respect to $K$. However, the GLSR algorithm achieves a high estimation accuracy of the delay parameters, as shown in Fig. 5. For example, when $K = 5$, both GLSR-1 and GLSR-2 yield a RRMS-TDE below 0.05, which is only about one-tenth of that obtained by OMP-1 and OMP-2. The RRMS-TDE of the GLSR algorithms increases as $K$ increases. On the other hand, the RRMS-TDE of the OMP-based algorithms maintains a high and nearly constant level, implying the modeling error between the assumed dictionary and the actual signals to be the dominant factor limiting the accuracy of time delay estimation. The proposed GLSR algorithm is more efficient to estimate the off-grid delays.

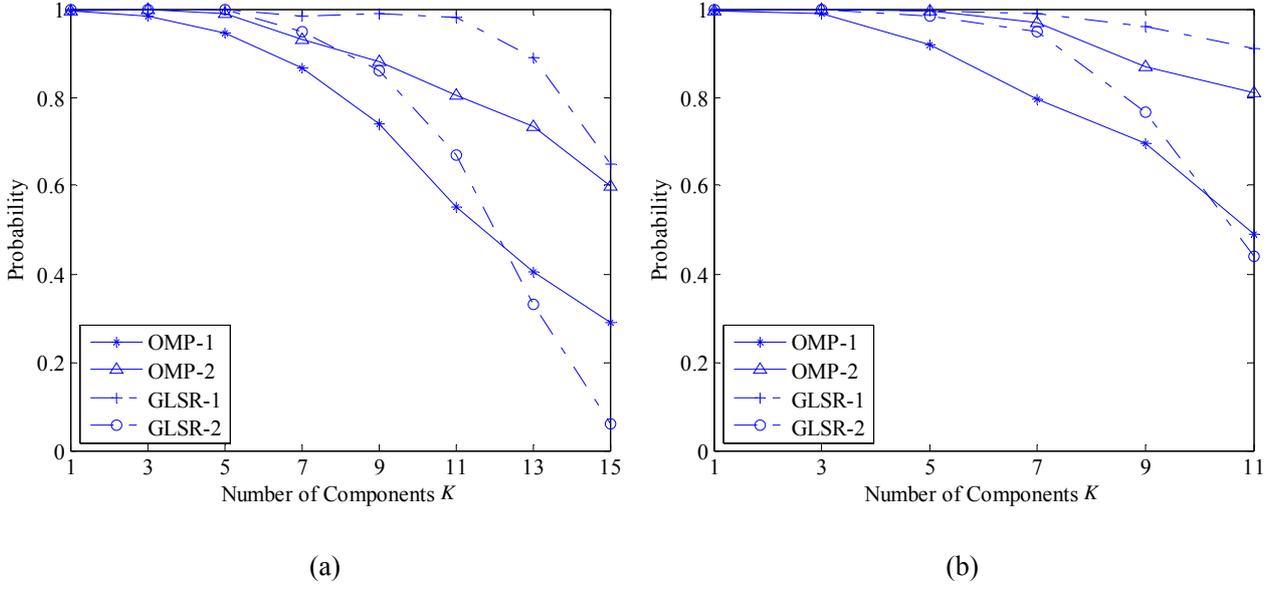

Fig.3 The probability of successful time-delay estimation versus the number of components $K$. (a) $B_{cs}=12.5\text{ MHz}$; (b) $B_{cs}=10\text{ MHz}$.

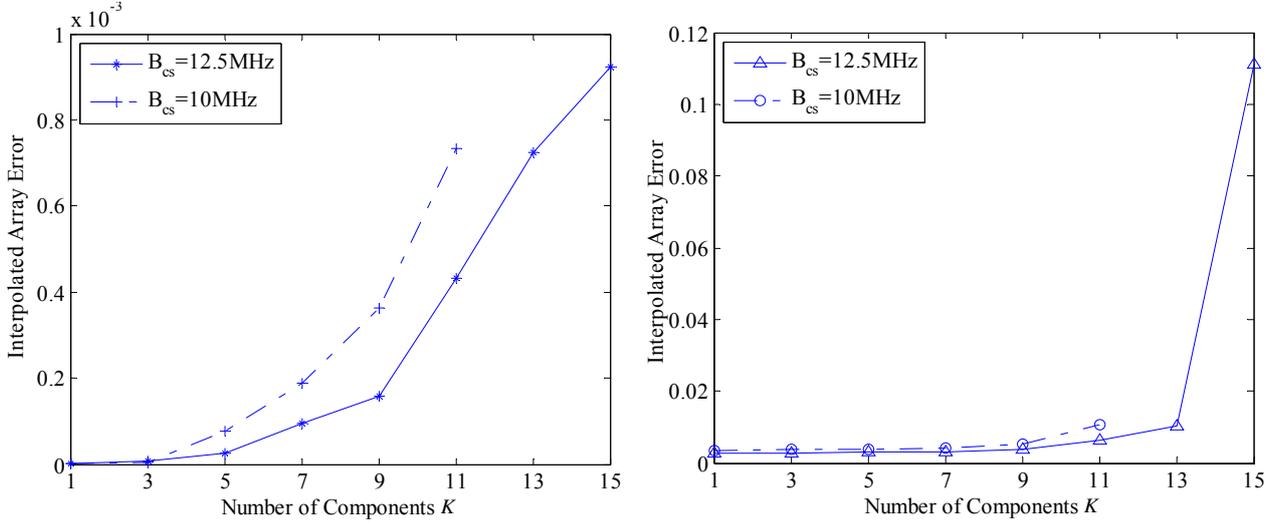

Fig.4 The interpolated array errors versus the number of components $K$. (a) GLSR-1; (b) GLSR-2.

## B. Reconstruction Performance in the Noise-Free Case

We define the relative root-mean-square signal reconstruction error (RRMS-SR), defined as

$$\text{RRMS-SR} = \frac{\left(\int_{-B/2}^{B/2}\left|\hat{S}(f)-\hat{\boldsymbol{\psi}}(f,\overline{\boldsymbol{\tau}})\overline{\tilde{\mathbf{v}}}\right|^{2}df\right)^{1/2}}{\left(\int_{-B/2}^{B/2}\left|\hat{S}(f)\right|^{2}df\right)^{1/2}}, \quad (61)$$

to evaluate the accuracy of signal reconstruction. Fig. 6 shows RRMS-SR versus the number of components $K$. It is observed that GLSR algorithms achieve much lower RRMS-SR than OMP-based algorithms when $K \leq 11$. For $B_{cs} = 12.5$ MHz, as $K$ increases, the RRMS-SR of GLSR algorithms increases fast than that of OMP-based algorithms. This is again due to the increases of the interpolated array error. Fig. 7 depicts the RRMS-SR performance versus the bandwidth $B_{cs}$. For $K = 5$ (Fig. 7(a)), the GLSR algorithm achieves much lower RRMS-SR than OMP-based recovery. For $K = 10$ (Fig. 7(b)), the RRMS-SR of GLSR-1 still offers the lowest errors, and GLSR-2 outperforms both OMP-1 and OMP-2 when the bandwidth $B_{cs}$ is larger than 13 MHz. As shown in Fig. 7, the increases of the bandwidth $B_{cs}$ will improve the RRMS-SR performance of GLSR algorithms. The reason is that, as the bandwidth $B_{cs}$ increases, we can collect more measurements and thus form a larger array in the proposed GLSR algorithm.

In comparisons of two GLSR algorithms, the GLSR-1 performs better than the GLSR-2 because of the accurate assumption of the sector $\Theta$, which leads to lower interpolated array error and better signal recovery performance.

It is also noted that OMP-2 performs better than OMP-1. However, because of highly coherent dictionary in OMP-2, there may be the instability in the signal recovery and lead to a large recovery error in some cases as shown in Fig.8. We removed these outliers when generating Figs. 6 and 7.

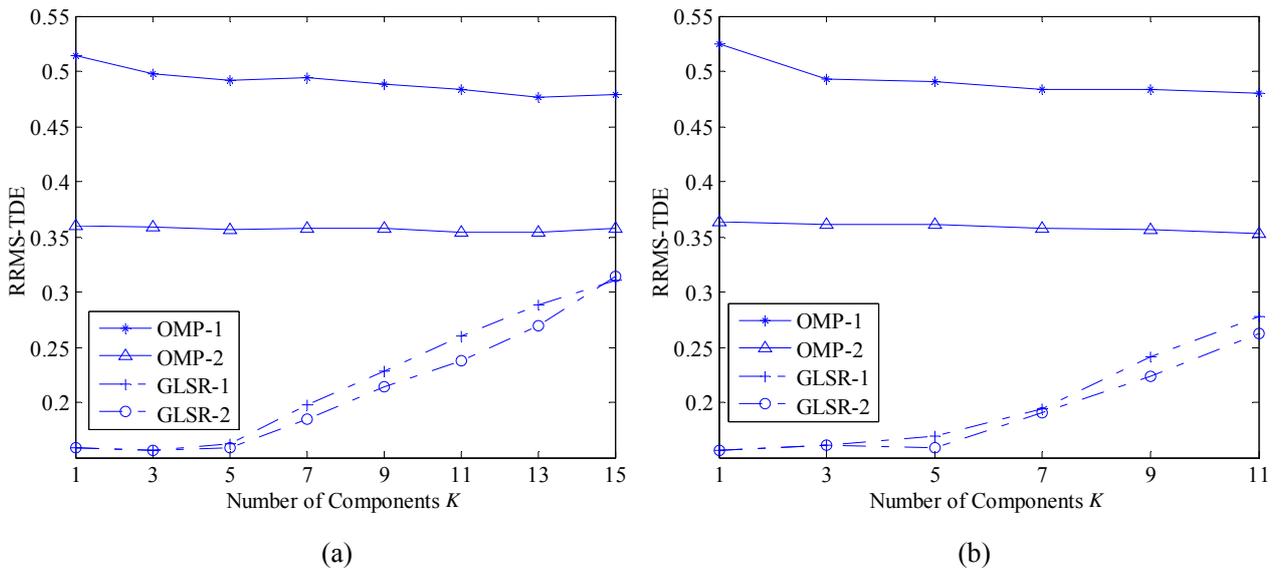

Fig. 5 The RRMS-TDE of the successful time-delay estimation versus the number of components $K$. (a) $B_{cs} = 12.5$ MHz; (b) $B_{cs} = 10$ MHz.

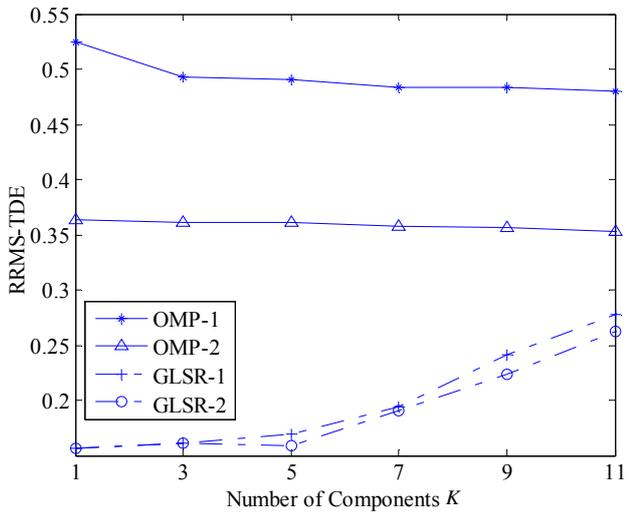
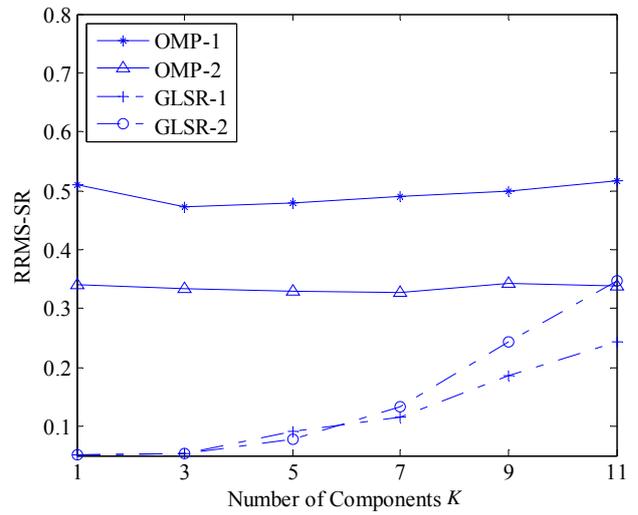

(a)                                   (b)

Fig. 6 The RRMS-SR versus the number of components $K$. (a) $B_{cs}=12.5MHz$; (b) $B_{cs}=10MHz$.

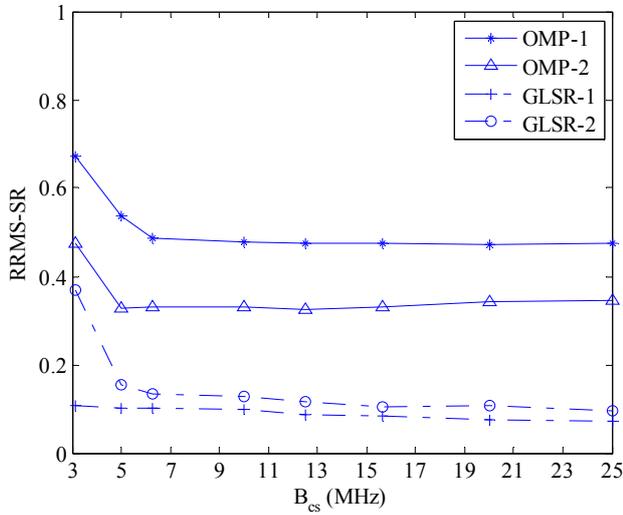
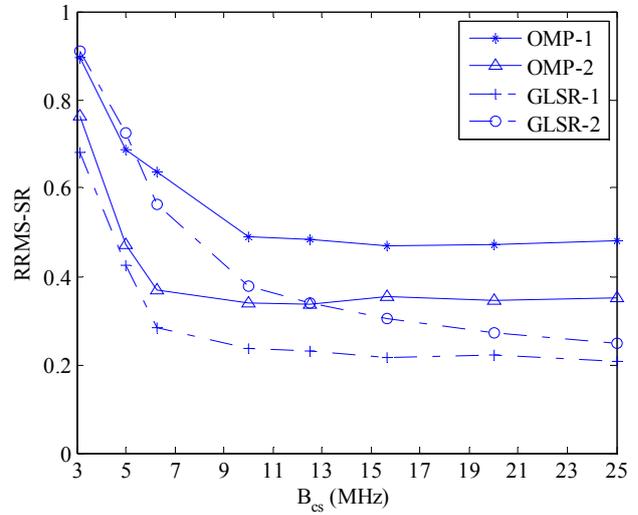

(a)                                   (b)

Fig.7 The RRMS-SR versus the bandwidth $B_{cs}$. (a) $K=5$; (b) $K=10$.

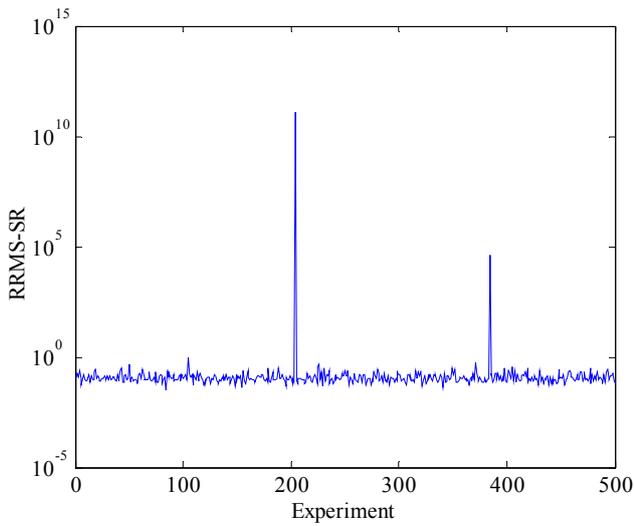
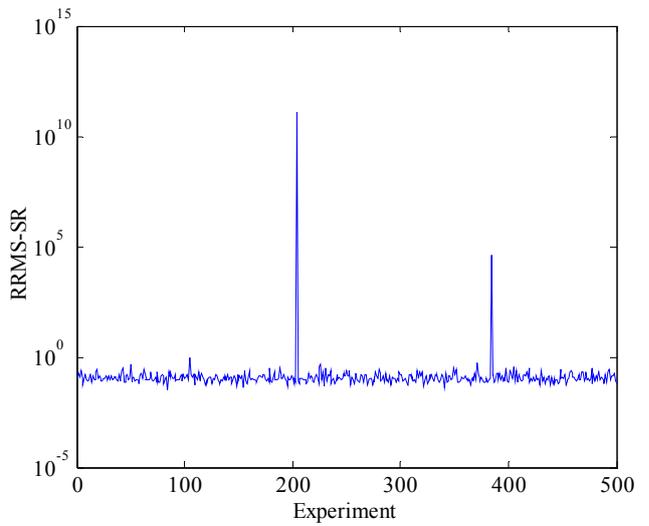

(a)                                   (b)

Fig.8 The RRMS-SR of OMP-2 algorithm in 500 Monte Carlo experiments with $K=15$. (a) $B_{cs}=12.5MHz$; (b) $B_{cs}=10MHz$.

*C. Reconstruction Performance in the Noisy Case*

In noisy scenarios, the IF signal is corrupted by a bandlimited additive white Gaussian noise with power spectral density $N_0/2$ spanning the bandwidth of $B$ centered at frequency $f_0$. The input SNR (ISNR) and the reconstructed SNR (RSNR), respectively defined as

$$\text{ISNR} = \frac{\int_{-B/2}^{B/2} |\hat{S}(f)|^2 df}{N_0 B}, \tag{62}$$

$$\text{RSNR} = \frac{\int_{-B/2}^{B/2} |\hat{S}(f)|^2 df}{\mathrm{E}\left[\int_{-B/2}^{B/2} |\hat{S}(f) - \hat{\psi}(f,\bar{\tau})\bar{\tilde{\mathbf{v}}}|^2 df\right]}, \tag{63}$$

are used to quantify the noise impact on the signal reconstruction.

The RSNR performance versus the ISNR is shown in Fig. 9. In both Figs. 9(a) and 9(b), the RSNR of OMP-based signal reconstruction is almost constant at a low level (respectively 7dB and 10dB for OMP-1 and OMP-2) as the ISNR increases from 10dB to 30dB. It implies that the effect of the grid mismatch on the signal reconstruction performance is much larger than that of the input noise. The RSNR of GLSR-1 and GLSR-2 increases almost linearly as the ISNR increases. For $K=5$, as shown in Fig. 9(a), GLSR-1 and GLSR-2 achieve about 8 to 10 dB SNR improvement as the ISNR varies from 10 dB to 30 dB. For $K=10$, as shown in Fig.9(b), the RSNR of GLSR algorithm does not improve as much due to a higher array interpolation error. However, at a high ISNR, both GLSR-1 and GLSR-2 outperform OMP-1 and OMP-2. The results show that the proposed GLSR algorithm is more robust to the noise than the OMP-based algorithm for a small $K$ and/or a high ISNR.

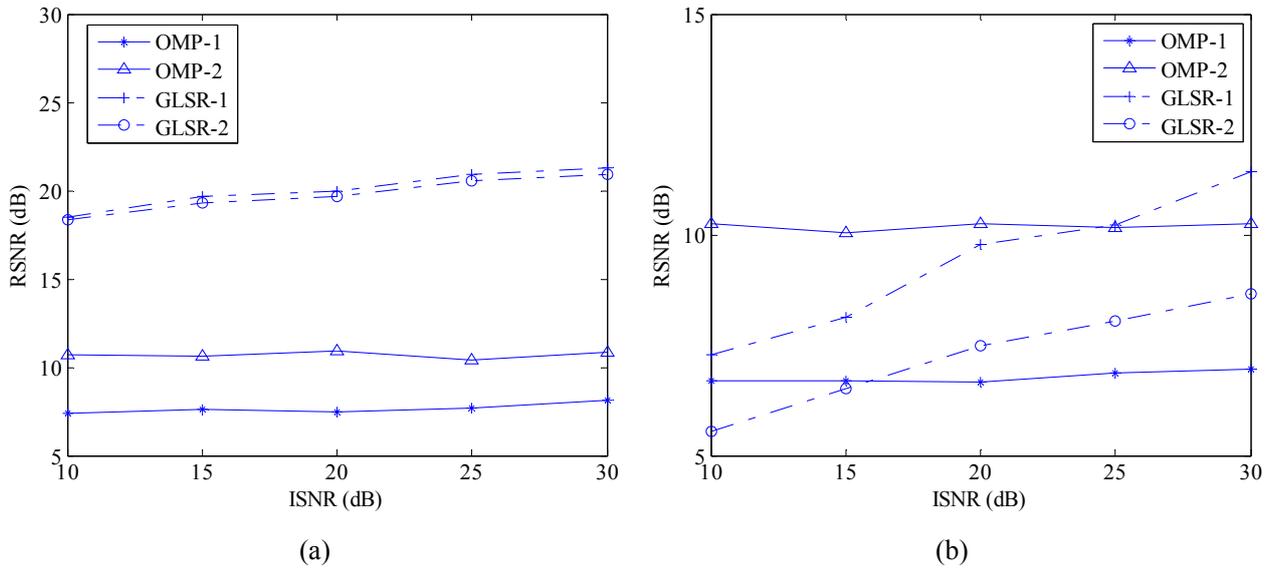

Fig. 9 The RSNR versus the ISNR for $B_{cs} = 12.5$ MHz. (a) $K = 5$; (b) $K = 10$.

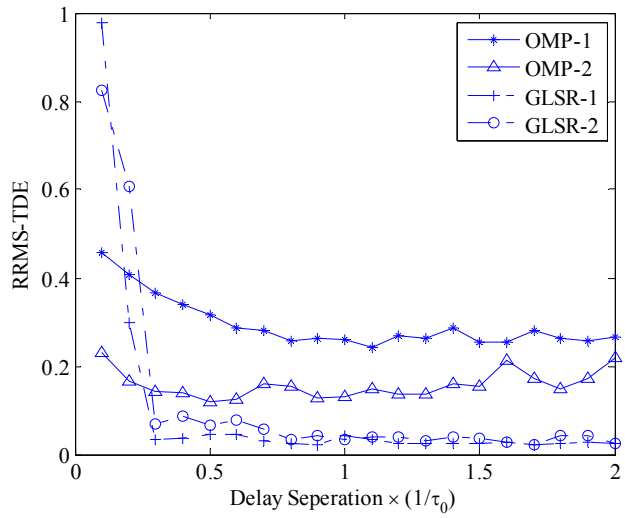

Fig. 10 The RRMS-TDE versus the delay separation for $B_{cs} = 12.5$ MHz.

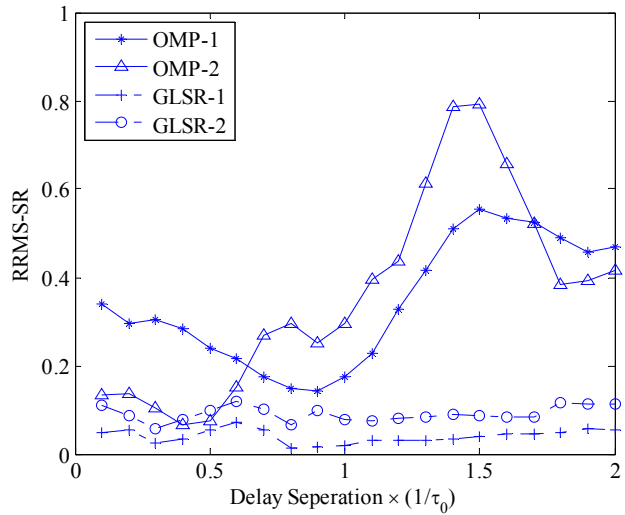

Fig. 11 The RRMS-SR versus the delay separation for $B_{cs} = 12.5$ MHz.

## D. Resolution Performance

We now compare the resolution performance to separate two closely located components between the OMP-based algorithm and the GLSR algorithm. The delay of one component is randomly chosen from a uniform distribution, and the other one is set apart between $0.1\tau_0$ to $2\tau_0$ with a uniform distribution. The gain coefficient is set to be fixed so that weaker components will not be masked by stronger ones.

In Fig. 10, we show the RRMS-TDE as delay separation between the two time-delay components. It is seen that the GLSR algorithm can achieve a much higher resolution when the delay separation is larger than $0.3\tau_0$. When the delay separation is smaller than $0.3\tau_0$, the GLSR algorithm identifies the two closely located components as a single component and thus leads to a large RRMS-TDE. In Fig. 11, the RRMS-SR versus the delay separation is shown. It is interesting to note that the GLSR algorithm maintains a low RRMS-SR even when the delay separation is smaller than $0.3\tau_0$ because, in this case, the combined waveform of the two components is similar to a single component with a combined gain coefficient. On the other hand, the RRMS-SR of the OMP-based algorithm changes dramatically as the delay separation increases. As such, it is demonstrated that the dictionary defined on the discretized grids cannot represent the two closely located components effectively.

## VIII. Discussion and Conclusion

As a new RF receiver architecture that is capable of capturing sparse wideband signals, the QuadCS system has wide applications in radar, communication and navigation systems. This paper makes important improvements from our earlier work [13] for application in practical environments where the delays are generally defined in a continuous space and cannot be characterized by a pre-designed grid structure. An effective and efficient solution for the estimation of delay parameters and gain coefficients in such scenarios is developed. The main contributions are summarized as follows:

(1) A parametric spectrum-matched dictionary is defined which well describes the received signals in radar and communications.

(2) An equivalent beamspace DOA formulation is developed for estimating delay parameters and gain coefficients. Sufficient conditions to guarantee the successful estimation of the delay parameters are theoretically derived.

(3) An interpolated array equivalence is established to utilize beamspace DOA estimation formulation with time-varying beamforming matrices.

(4) Extensive simulation results verify that the proposed GLSR algorithm achieves a super-resolution estimation with a high recovery accuracy.

We maintain that the introduced DOA-based technique for solving the off-grid problems can be applied to other analog-to-information conversion problems provided that the sampling sequences can be decoupled as described in (27) and (30). The proposed GLSR algorithm works well when the signal components exhibit a low sparsity. The performance gradually degrades with the signal sparsity as it suffers from a higher interpolated array errors. Reducing the interpolated array error deserves further research.

REFERENCES

[1] K. C. Ho, Y. T. Chan, and R. Inkol, "A digital quadrature demodulation system," IEEE Trans. Aerospace and Electronic Systems, vol. 32, no. 4, pp. 1218-1227, Oct. 1996.

[2] R. G. Vaughan, N. L. Scott, and D. R. White, "The theory of bandpass sampling," IEEE Trans. Signal Process., vol. 39, no. 9, pp. 1973-1984, Sep. 1991.

[3] E. Candès and T. Tao, "Decoding by linear programming," IEEE Trans. Inf. Theory, vol. 51, no. 12, pp. 4203-4215, Dec. 2005.

[4] D. Donoho, "Compressed sensing," IEEE Trans. Inf. Theory, vol. 52, no. 4, pp. 1289-1306, Apr. 2006.

[5] E. Candès, J. Romberg, and T. Tao, "Robust uncertainty principles: exact signal reconstruction from highly incomplete frequency information," IEEE Trans. Inf. Theory, vol. 52, no. 2, pp. 489-509, Feb. 2006.

[6] J. Laska, S. Kirolos, Y. Massoud, R. Baraniuk, A. Gilbert, M. Iwen, and M. Strauss,"Random sampling for analog-to-information conversion of wideband signals," in Proc. IEEE Dallas Circuits and Systems Workshop (DCAS), Dallas, TX, Oct. 2006, pp.119-122.

[7] J. Tropp, M. Wakin, M. Duarte, D. Baron, and R. Baraniuk,"Random filters for compressive sampling and reconstruction," in Proc. IEEE Int. Conf. on Acoustics, Speech, and Signal Processing (ICASSP), Toulouse, France, May 2006.

[8] J. Laska, S. Kirolos, M. Duarte, T. Ragheb, R. Baraniuk, and Y. Massoud,"Theory and implementation of an analog-to-information converter using random demodulation," in Proc. IEEE Int. Symp. on Circuits and Systems (ISCAS), New Orleans, LO, May 2007, pp. 1959-1962.

[9] J. A. Tropp, J. N. Laska, M. F. Duarte, J. K. Romberg, and R. G. Baraniuk,"Beyond Nyquist: efficient sampling of sparse bandlimited signals," IEEE Trans. Inf. Theory, vol. 56, no. 1, pp.520-544, Jan. 2010.

[10] O. Taheri and S. A. Vorobyov, "Segmented compressed sampling for analog-to-information conversion: method and performance analysis," IEEE Trans. Signal Process., vol. 59, no. 2, pp. 554-572, Feb. 2011.

[11] M. Mishali, Y. C. Eldar, and A. Elron, "Xampling: signal acquisition and processing in union of subspaces," IEEE Trans. Signal Process., vol. 59, no. 10, pp. 4719-4734, Oct. 2011.

[12] F. Xi, S-Y. Chen, and Z. Liu, "Quadrature compressive sampling for radar echo signals," in Proc. Int. Conf. on Wireless Commun. and Signal Process. (WCSP), Nanjing, China, Nov. 2011.

[13] F. Xi, S-Y. Chen, and Z. Liu, "Quadrature compressive sampling for radar signals," IEEE Trans. Signal Process., vol. 62, no. 11, pp.2787-2802, Jun. 2014.

[14] F. Xi, S-Y. Chen, and Z. Liu, "Quadrature compressive sampling for radar signals: output noise and robust reconstruction," in Proc. IEEE China Summit & Int. Conf. on Signal and Inf. Process. (ChinaSIP), Xi'an, China, Jul. 2014, pp. 790-794.

[15] C. Liu, F. Xi, S-Y. Chen, Y. D. Zhang, and Z. Liu, "Pulse-doppler signal processing with quadrature compressive sampling," IEEE Trans. Aerospace and Electronic Systems, accept for publication.

[16] G. L. Turin, "Introduction to spread-spectrum antimultipath techniques and their application to unban digital radio," Proc. IEEE, vol. 68, no. 3, pp. 328-353, Mar. 1980.

[17] J. Soubielle, I. Fijalkow, P. Duvaut, and A. Bibaut, "GPS positioning in a multipath environmrnt," IEEE Trans. Signal Process., vol. 50, no. 1, pp.141-150, Jan. 2002.

[18] G. Shi, J. Lin, X. Chen, F. Qi, D. Liu, and L. Zhang, "UWB echo signal detection with ultra-low rate sampling based on compressed sensing," IEEE Trans. Circuits and Systems II: Express Briefs, vol. 55, no. 4, pp. 379-383, Apr. 2008.

[19] Y. Chi, L. Scharf, A. Pezeshki, and A. Calderbank, "Sensitivity to basis mismatch in compressed sensing," IEEE Trans. Signal Process., vol. 59, no. 5, pp.2182-2195, May 2011.

[20] A. Fannjiang and W. Liao, "Coherence pattern-guided compressive sensing with unresolved grids," SIAM J. Imaging Sciences, vol. 5, no. 1, pp. 179-202, Jan. 2012.

[21] M. F. Duarte and R. G. Baraniuk, "Spectral compressive sensing," Applied and computational Harmonic Analysis, vol. 35, no. 1, pp. 111-129, Jul. 2013.

[22] H. Zhu, G. Leus, and G. Giannakis, "Sparsity-cognizant total least-squares for perturbed compressive sampling," IEEE Trans. Signal Process., vol. 59, no. 5, pp. 2002-2016, May 2011.

[23] Z. Yang, L. Xie, and C. Zhang, "Off-grid direction of arrival estimation using sparse Bayesian inference," IEEE Trans. Signal Process., vol.61, no.1, pp.38-43, Jan. 2013.

[24] C. Ekanadham, D. Tranchina, and E. P. Simoncelli, "Recovery of sparse translation-invariant signals with continuous basis pursuit," IEEE Trans. Signal Process., vol. 59, no. 10, pp. 4735-4744, Oct. 2011.

[25] L. Hu, Z. Shi, J. Zhou, and Q. Fu, "Compressed sensing of complex sinusoids: An approach based on dictionary refinement," IEEE Trans. Signal Process., vol. 60, no. 7, pp. 3809-3822, Jul. 2012.

[26] J. Fang, J. Li, Y. Shen, H. Li, and S. Li, "Super-resolution compressed sensing: An iterative reweighted algorithm for joint parameter learning and sparse signal recovery," IEEE Signal Process. Lett., vol. 21, no. 6, pp. 761-765, Jun. 2014.

[27] Z. Tan, P. Yang, A. Nehorai, "Joint sparse recovery method for compressed sensing with structured dictionary mismatches," IEEE Trans. Signal Process., vol. 62, no. 19, pp. 4997-5008, Oct. 2014.

[28] E. Candes and C. Fernandez-Granda, "Towards a mathematical theory of super-resolution," Commun. Pure Appl. Math., vol. 67, no. 6, pp. 906-956, Jun. 2014.

[29] G. Tang, B. N. Bhaskar, P. Shah, and B. Recht, "Compressed sensing off the grid," IEEE Trans. Inf. Theory, vol. 59, no. 11, pp. 7465-7490, Nov. 2013.

[30] Y. Chen and Y. Chi, "Robust spectral compressed sensing via structured matrix completion," IEEE Trans. Inf. Theory, vol. 60, no. 10, pp. 6576-6601, Oct. 2014.

[31] S. Ji, Y. Xue, and L. Carin, "Bayesian compressive sensing," IEEE Trans. Signal Process., vol. 56, no. 6, pp. 2346-2356, Jun. 2008.

[32] S. Babacan, R. Molina, and A. Katsaggelos, "Bayesian compressive sensing using Laplace priors," IEEE Trans. Image Process., vol. 19, no. 1, pp. 53-63, Jan. 2010.

[33] V. Chandrasekaran, B. Recht, P. A. Parrilo, and A. S. Willsky, "The convex geometry of linear inverse problems," Foundations of Computational Mathematics, vol. 12, no. 6, pp. 805-849, Oct. 2012.

[34] R. Schmidt, "Multiple emitter location and signal parameter estimation," IEEE Trans. Antennas Propag., vol. AP-34, no. 3, pp. 276-280, Mar. 1986.

[35] R. Roy and T. Kailath, "ESPRIT-estimation of signal parameters via rotational invariance techniques," IEEE Trans. Acoust., Speech, Signal Process., vol. 37, no. 7, pp. 984-995, Jul. 1989.

[36] P. Stoica and A. Nehorai, "Comparative performance study of element-space and beam-space MUSIC estimators," Circuits Syst. Signal Process., vol. 10, no. 3, pp. 285-292, 1991.

[37] X. L. Xu and K. M. Buckley, "An analysis of beamspace source localization," IEEE Trans. Signal Process., vol. 41, no. 1, pp. 501-504, Jan. 1993.

[38] H. B. Lee and M. S. Wengrovitz, "Resolution threshold of beamspace MUSIC for two closely spaced emitters," IEEE Trans. Acoust. Speech Signal Process., vol. 38, no. 9, pp. 1545-1559, Sep. 1990.

[39] B. Friedlander and A. Weiss, "Direction finding using spatial smoothing with interpolated arrays," IEEE Trans. Aerospace and Electronic Systems, vol. 28, no. 2, pp. 574-587, Apr. 1992.

[40] B. Friedlander and A. Weiss, "Direction finding for wideband signals using an interpolated array," IEEE Trans. Signal Process., vol. 41, no. 4, pp.1618-1634, Apr. 1993.

[41] A. Weiss, B. Friedlander, and P. Stoica, "Direction-of-arrival estimation using MODE with interpolated arrays," IEEE Trans. Signal Process., vol. 43, no. 1, pp.296-300, Jan. 1995.

[42] K. Gedalyahu and Y. C. Eldar, "Time-delay estimation from low-rate samples: a union of subspaces approach," IEEE Trans. Signal Process., vol. 58, no. 6, pp. 3017-3031, Jun. 2010.

[43] Y. M. Lu and M. N. Do, "A theory for sampling signals from a union of subspaces," IEEE Trans. Signal Process., vol. 56, no. 6, pp. 2334-2345, Jun. 2008.

[44] R. Tibshirani, "Regression shrinkage and selection via the lasso," J. Royal. Statist. Soc B., vol. 58, no. 1, pp. 267-288, 1996.

[45] J. Haupt, W. Bajwa, G. Raz, and R. Nowak, "Toeplitz compressed sensing matrices with applications to sparse channel estimation," IEEE Trans. Inf. Theory, vol. 56, no. 11, pp. 5862-5875, Nov. 2010.

[46] A. Zeira and B. Friedlander, "On the performance of direction finding with time-varying arrays," Signal Processing, vol. 43, pp. 133-147, May 1995.

[47] B. Friedlander and A. Zeira, "Eigenstructure-based algorithms for direction finding with time-varying array," IEEE Aerospace and Electronic Systems, vol. 32, no. 2, pp. 689-701, Apr. 1996.

[48] M. Wax and I. Ziskind, "On unique localization of multiple sources by passive sensor arrays," IEEE Trans. Acoust., Speech, Signal Process., vol. 37, no. 7, pp. 996-1000, Jul. 1989.

[49] M. Wax and T. Kailath, "Detection of signals by information theoretic criteria," IEEE Trans. Acoustics, Speech, Signal Process., vol. 33, no. 4, pp. 387-392, Apr. 1985.

[50] J. A. Tropp and A. C. Gilbert, "Signal reocvery from random measurements via orthogonal matching pursuit," IEEE Trans. Signal Process., vol. 53, no. 12, pp.4655-4666, Dec. 2007.